\documentclass[a4paper,fleqn]{cas-dc}

\usepackage[numbers]{natbib}
\usepackage{lineno}

\def\tsc#1{\csdef{#1}{\textsc{\lowercase{#1}}\xspace}}
\tsc{WGM}
\tsc{QE}
\tsc{EP}
\tsc{PMS}
\tsc{BEC}
\tsc{DE}

\begin{document}
\let\WriteBookmarks\relax
\def\floatpagepagefraction{1}
\def\textpagefraction{.001}

\shorttitle{Stochastic dose fluctuations in boron neutron capture therapy with heterogeneous boron distributions}
\shortauthors{R.~Fardela, R.~Muslim, T.~A.~I.~Kusuma }

\title[mode = title]{Stochastic dose fluctuations in boron neutron capture therapy with heterogeneous boron distributions}    

\author[1]{Ramacos Fardela} [orcid=0000-0002-6991-4156]
\ead{ramacosfardela@sci.unand.ac.id}
\cormark[1]
\cortext[cor1]{Corresponding author}

\author[2, 3, 4]{Roni Muslim}[orcid= 0000-0001-6925-5923]
\ead{roni.muslim@apctp.org}

\author[5]{Tista Artu Indra Kusuma}
\ead{tista@rs-jih.co.id}

\affiliation[1]{
    organization={Department of Physics, Faculty of Mathematics and Natural Science, Universitas Andalas},
    state={Sumatera Barat},
    postcode={25163},
    country={Indonesia}
}

\affiliation[2]{
    organization={Asia Pacific Center for Theoretical Physics (APCTP)},
    city={Pohang},
    postcode={37673},
    country={Republic of Korea}
}
    
\affiliation[3]{organization={Research Center for Quantum Physics,
    National Research and Innovation Agency (BRIN)},
    city={South Tangerang},
    postcode={15314}, 
    country={Indonesia}}
    
\affiliation[4]{organization={Indonesian Nuclear Technology Polytechnic,
    National Research and Innovation Agency (BRIN)},
    city={Yogyakarta},
    postcode={55281}, 
    country={Indonesia}}

\affiliation[5]{
    organization={JIH Hospital Yogyakarta},
    state={Yogyakarta},
    postcode={55283},
    country={Indonesia}
}

\begin{abstract}
Dose fluctuations at the cellular scale in
boron neutron capture therapy (BNCT) arise
from stochastic neutron capture and spatially
heterogeneous boron distributions.
This study develops a stochastic framework
combining energy-deposition responses from
Geant4 simulations, a Poisson capture
process, and spatially correlated lognormal
boron fields to investigate nuclear dose
variance and zero-hit probability.
The results show that the variance contribution
from stochastic neutron capture increases
linearly with neutron fluence, whereas the
contribution from boron heterogeneity increases
quadratically. These distinct scaling
behaviors produce a correlation-length-dependent
crossover between the dominant sources
of dose fluctuations.
For a heterogeneity amplitude of $H=0.8$
and a correlation length of
$\xi=3\,\mu\mathrm{m}$, the crossover occurs
at approximately
$1.50\times10^{12}\,\mathrm{cm}^{-2}$,
while the ensemble relative dose variance
approaches a plateau of $0.160$ at high
fluence. Boron heterogeneity also increases
the probability that a nucleus receives
no energy deposition from alpha--lithium
reaction products. At $\xi=6\,\mu\mathrm{m}$
and $\Phi_n=3\times10^{11}\,\mathrm{cm}^{-2}$,
the zero-hit probability is $40.49\%$
higher than that for a homogeneous boron
distribution with the same mean concentration.
This study shows that spatial
boron correlations and stochastic neutron
capture influence nuclear dose fluctuations
and zero-hit statistics beyond what can
be characterized by mean boron
concentration alone.
\end{abstract}






\maketitle

\section{Introduction}
\label{sec:introduction}

Boron neutron capture therapy (BNCT) is a
radiotherapy approach that combines the
delivery of boron-containing compounds
with neutron irradiation. Neutron capture
by $^{10}$B produces an alpha particle
and a $^{7}$Li nucleus, which deposit
their energy over distances of a few
micrometers, comparable to the dimensions
of cells and their nuclei. The short
range of these particles provides the
basis for the potential selectivity of
BNCT, but also makes energy deposition
within cellular structures highly
dependent on the location of boron
and the sites of neutron capture
\cite{Barth2005,Dymova2020}.
Advances in BNCT technology, particularly
accelerator-based neutron sources,
boron delivery agents, and imaging
techniques, have expanded its potential
applications in cancer treatment
\cite{Barth2012,BarthWu2024,Zheng2026}.
Clinical progress, especially in the
treatment of head and neck cancers,
has further highlighted the need for
dose calculations that account not
only for energy deposition at the
tissue level but also for its
distribution at the cellular scale
\cite{Hirose2021,Shen2024,ZhouClin2025}.

One of the main challenges in BNCT
dosimetry is the heterogeneous
distribution of boron. Delivery agents
such as boronophenylalanine (BPA) and
sodium borocaptate (BSH) can produce
different boron concentrations across
tissues, individual cells, and
subcellular compartments
\cite{MontiHughes2023,BarthGupta2024}.
New boron delivery agents continue
to be developed to improve boron
accumulation and retention in
target tissues
\cite{Xu2024,Yang2026,SelgDelivery2026}.
Nevertheless, tumor heterogeneity
and variations in boron localization
remain important challenges when
relating mean boron concentration
to the dose received by cellular
structures \cite{ZhouHetero2025}.
The influence of nonuniform boron
distributions has been demonstrated
in dose calculations based on
$^{18}$F-BPA positron emission
tomography (PET), which show
differences between heterogeneous
and homogeneous boron assumptions
\cite{Teng2023,Kobayashi2025}.
Recent PET-based modeling studies
have also emphasized the importance
of representing boron heterogeneity
in treatment planning and in
evaluating dose--toxicity relationships
\cite{TengHypo2026,TengMucosal2026}.
However, tissue-level imaging cannot
fully resolve variations in boron
concentration at the subcellular scale.
Developments in neutron autoradiography,
particle track detectors, and boron
isotope quantification have further
highlighted the importance of
characterizing boron microdistributions
\cite{Wu2025BoronAutoradiography,
Galuzzi2025FNTD,Selg2026BoronQuantification}.
Beyond the magnitude of concentration
variations, the spatial arrangement
of boron is also important.
Two boron distributions with the
same mean concentration and
heterogeneity amplitude may have
different correlation lengths,
leading to different degrees of
spatial averaging as the reaction
products deposit energy within
the nucleus.

The effects of boron microdistribution
and stochastic energy deposition
have been investigated through
several approaches to BNCT
microdosimetry. Santa Cruz and
Zamenhof developed an analytical
framework that accounts for boron
localization and stochastic energy
deposition at microscopic scales
\cite{SantaCruzZamenhof2004}.
Sato \emph{et al.} subsequently
developed stochastic microdosimetric
kinetic models incorporating
intra- and intercellular boron
heterogeneity to estimate biological
effectiveness \cite{Sato2018,Sato2021}.
Particle transport simulations
have also been used to investigate
energy deposition in microscopic
targets \cite{Han2023}, while
microdosimetric measurements at
different site sizes have
demonstrated the importance of
spatial scale in characterizing
energy deposition \cite{Selva2024}.
More recently, cellular-scale
simulation approaches have been
developed to relate energy-deposition
statistics to biological response
parameters in BNCT
\cite{Shigehira2026,Yamazaki2026}.
Monte Carlo (MC) transport methods have
also continued to advance for
dose calculation and treatment
planning \cite{Zhong2023,Pistone2025}.
Other recent studies have combined
boron microdistribution information
with dosimetric modeling, including
stochastic representations of
boron concentrations in cellular
compartments
\cite{Lin2025BoronCBE,Marcaccio2026}.
Together, these developments show
that mean absorbed dose alone may
not adequately describe the
stochastic nature of energy
deposition at microscopic scales
\cite{SantaCruz2026}.

Although the importance of boron
heterogeneity is well recognized,
the relationship between its spatial
correlation length and the relative
contributions of different mechanisms
to dose fluctuations requires
further characterization.
For a fixed boron distribution,
dose fluctuations arise from
the random number of neutron
capture events and the energy
deposited by each event.
In contrast, variations between
boron distributions produce
differences in conditional mean
dose that depend on their spatial
correlation structure.
These two mechanisms have different
dependencies on neutron fluence,
so their relative contributions
can change even when the mean
boron concentration remains fixed.
Spatial correlations also affect
the probability that a nucleus
receives no energy deposition
from the reaction products.
A statistical framework is therefore
needed to connect boron correlation
length with dose variance
decomposition and zero-hit
probability.

In this study, the boron distribution
is represented by a spatially
correlated lognormal random field,
while neutron capture events are
modeled as a Poisson process
conditional on the boron field.
Variations in capture intensity
between field realizations give
rise to a Cox process
\cite{Moller1998}.
Single-event energy-deposition
responses, including the mean
deposited energy, its second
raw moment, and the hit
probability, are obtained from
Geant4 simulations
\cite{Agostinelli2003,Allison2016}
and combined with the stochastic
formulation to determine nuclear
dose statistics. This approach
allows the total dose variance
to be separated into contributions
from stochastic neutron capture
and spatial boron heterogeneity,
with their dependence on neutron
fluence determined analytically.
The correlation length is varied
while keeping the mean boron
concentration and heterogeneity
amplitude fixed to determine
the fluence at which the two
variance contributions become
equal and the limiting relative
variance at high fluence.
The zero-hit probability is also
calculated and compared with
the homogeneous-boron reference.
Analytical predictions are compared
with MC sampling to
assess the consistency of the
formulation and its numerical
implementation.

The contribution of this
study is a quantitative
characterization of how boron
correlation length influences
the mechanisms of dose fluctuations
and the zero-hit probability
at the nuclear scale.
Unlike approaches that directly
predict biological responses,
the present work focuses on
the physical statistics of
energy deposition by
alpha--lithium reaction products.
The analysis is restricted to
a spherical nucleus in a
water-equivalent medium with
a spatially uniform neutron
field and electromagnetic
transport of the reaction
products. The zero-hit
probability is not interpreted
as a cell survival probability,
since other BNCT radiation
components and cellular
biological responses are
not included in the model.

\section{Model and Methods}
\label{sec:model}

\subsection{Stochastic boron deposition}
\label{subsec:stochastic_model}

The model describes how spatial boron heterogeneity
influences the stochastic nature of neutron capture
and energy deposition in cellular targets, as
previously investigated in BNCT microdosimetry
\cite{SantaCruzZamenhof2004,Sato2018,Marcaccio2026}.
The surrounding medium is assumed to be
water-equivalent, with fixed composition and density,
while the distribution of $^{10}$B is allowed to vary
spatially. These variations are assumed not to
affect the transport properties of the medium
or the neutron field. The number density of
$^{10}$B atoms is represented by a spatially
correlated lognormal random field,
\begin{equation}
C_B(\mathbf r)
=
\overline C_B
\exp\left[sX(\mathbf r)-\frac{s^2}{2}\right],
\label{eq:boron_field}
\end{equation}
where $\overline C_B=\mathbb E[C_B]$ and
$X(\mathbf r)$ is a stationary Gaussian
random field with zero mean and unit variance.
The lognormal representation ensures that
the local boron concentration remains positive
and provides a random intensity field suitable
for a Cox process \cite{Moller1998}.
In the neutron capture equations, $C_B$
denotes the number of atoms per unit volume.
The mass concentration used as a numerical
input is converted into this quantity.

The Gaussian correlation function is chosen as
$\rho_X(d)=\exp[-d^2/(2a^2)]$, where
$d=|\mathbf r-\mathbf r'|$ and $a$ is the
Gaussian correlation-length parameter.
The heterogeneity amplitude is defined by
$H=\sigma_B/\overline C_B$, giving
$s^2=\ln(1+H^2)$. The boron concentration
covariance,
$K_C(d)=\operatorname{Cov}[C_B(\mathbf r),
C_B(\mathbf r')]$, and its normalized
correlation function are given by
\begin{equation}
\begin{aligned}
K_C(d)
&=
\overline C_B^{\,2}
\left\{\exp[s^2\rho_X(d)]-1\right\},
\\
\rho_B(d)
&=
\frac{K_C(d)}{\overline C_B^{\,2}H^2},
\end{aligned}
\label{eq:model_boron_covariance}
\end{equation}
for $H>0$. The boron correlation length
$\xi$ is defined through
$\rho_B(\xi)=\exp(-1/2)$.
This definition applies to the concentration
field after the lognormal transformation,
so $a$ and $\xi$ are generally different.
The Gaussian correlation parameter
satisfying this definition is obtained from
\begin{equation}
\begin{aligned}
\eta(H)
&=
\frac{\ln[1+H^2\exp(-1/2)]}{\ln(1+H^2)},
\\
a
&=
\frac{\xi}{\sqrt{-2\ln\eta(H)}}.
\end{aligned}
\label{eq:model_correlation_calibration}
\end{equation}
This construction allows the heterogeneity
amplitude and the correlation length of the
boron concentration field to be specified
independently. For $H=0$, the boron field
becomes homogeneous, and a correlation
length is no longer required.

For a given realization of the boron field,
neutron capture locations follow an
inhomogeneous Poisson process with the
time-integrated capture intensity
\begin{equation}
\lambda_C(\mathbf r)
=
F C_B(\mathbf r),
\label{eq:capture_intensity}
\end{equation}
where
$F=\int\sigma_{n\alpha}(E)\Phi_n(E)\,dE$
is a dimensionless capture factor.
Here, $\Phi_n(E)$ denotes the differential
neutron fluence integrated over the
irradiation period, and
$\sigma_{n\alpha}(E)$ is the cross section
for the $^{10}$B$(n,\alpha)^{7}$Li reaction.
Under the thermal-neutron approximation
used in the numerical calculations,
$F=\sigma_{\mathrm{th}}\Phi_n$.
The quantity $\lambda_C$ represents
the expected number of neutron captures
per unit volume. Accordingly, the number
of captures within a volume element
$\Delta V$, conditional on $C_B$, follows
\begin{equation}
N_{\mathrm{cap}}(\Delta V)\mid C_B
\sim
\operatorname{Poisson}
\left[
\int_{\Delta V}
\lambda_C(\mathbf r)\,d^3\mathbf r
\right].
\label{eq:poisson_capture}
\end{equation}
In the voxel-based discretization,
this integral is approximated by the
capture intensity at the voxel center
multiplied by its volume.
Capture counts in disjoint volumes
are independent when conditioned on
a fixed boron field \cite{LastPenrose2017}.
Correlations between different regions
arise when variations in the boron field
are included. The model neglects boron
depletion, changes in its spatial
distribution during irradiation,
and modifications of the neutron field
caused by boron heterogeneity.

Each neutron capture produces an
alpha--lithium pair, which is treated
as a single combined event.
The energy deposited by both reaction
products is summed within each history
to preserve the correlations associated
with their emission directions and
energy deposition. For a target $i$
with mass $m_i$ and source region
$\mathcal V$, the total deposited
energy and absorbed dose are defined as
\begin{equation}
\begin{aligned}
&E_i
=
\sum_{j=1}^{N_{\mathrm{cap}}(\mathcal V)}
\varepsilon_i(\mathbf r_j,\omega_j), \\
&D_{B,i}
=
\frac{E_i}{m_i},
\label{eq:cellular_dose}
\end{aligned}
\end{equation}
where $\mathbf r_j$ is the capture
location and $\omega_j$ includes
the reaction branch, emission
directions, and transport of the
reaction products and their secondaries.
The deposited energy $\varepsilon_i$
is zero when an event does not deposit
any energy in the target.
All histories, including those with
zero energy deposition, are included
when calculating the response moments,
\begin{equation}
\begin{aligned}
&u_i(\mathbf r)
=
\mathbb E_\omega[\varepsilon_i(\mathbf r,\omega)],\\
&v_i(\mathbf r)
=
\mathbb E_\omega[\varepsilon_i^2(\mathbf r,\omega)].
\label{eq:event_moments}
\end{aligned}
\end{equation}
Here, $v_i$ is the second raw moment,
whereas the variance of the energy
deposited by a single event at a fixed
location is $v_i-u_i^2$.
For a spherical target in a homogeneous
medium, rotational symmetry allows
the responses to be expressed as
functions of distance from the target
center, denoted by $u(r)$ and $v(r)$.
Deposited energy is converted into
joules and target mass into kilograms
when the absorbed dose is expressed
in gray.

\subsection{Nuclear dose statistics}
\label{subsec:analytical_theory}

Particle transport histories from
different capture events are assumed
to be conditionally independent for
a given boron field. By treating
the transport outcomes as marks
of the Poisson process, the moment
formulas for Poisson integrals give
\cite{LastPenrose2017}
\begin{align}
\mathbb E[D_{B,i}\mid C_B]
&=
\frac{1}{m_i}
\int_{\mathcal V}
\lambda_C(\mathbf r)
u_i(\mathbf r)\,d^3\mathbf r,
\label{eq:mean_dose}
\\
\operatorname{Var}(D_{B,i}\mid C_B)
&=
\frac{1}{m_i^2}
\int_{\mathcal V}
\lambda_C(\mathbf r)
v_i(\mathbf r)\,d^3\mathbf r.
\label{eq:conditional_variance}
\end{align}
The conditional mean dose is determined
by the mean energy-deposition response
weighted by the local capture intensity.
The conditional variance involves the
second raw moment because it accounts
for fluctuations in both the number
of capture events and the energy
deposited by each event.
The mean response $u_i$ alone is
therefore insufficient to determine
the dose variance.

Averaging over the ensemble of
boron fields gives
\begin{equation}
\langle D_{B,i}\rangle
=
\frac{F\overline C_B}{m_i}
\int_{\mathcal V}
u_i(\mathbf r)\,d^3\mathbf r.
\label{eq:model_ensemble_mean_dose}
\end{equation}
For the same mean boron concentration
and transport response, the ensemble
mean dose is identical to that obtained
under the homogeneous-boron assumption.
The effect of boron heterogeneity
instead appears in the dose
fluctuations. Applying the law
of total variance yields
\begin{equation}
\operatorname{Var}(D_{B,i})
=
\sigma_{\mathrm{shot},i}^2
+
\sigma_{\mathrm{het},i}^2,
\label{eq:variance_decomposition}
\end{equation}
where
\begin{align}
\sigma_{\mathrm{shot},i}^2
&=
\mathbb E_{C_B}
[\operatorname{Var}(D_{B,i}\mid C_B)]
\nonumber\\
&=
\frac{F\overline C_B}{m_i^2}
\int_{\mathcal V}
v_i(\mathbf r)\,d^3\mathbf r,
\label{eq:shot_variance}
\\
\sigma_{\mathrm{het},i}^2
&=
\operatorname{Var}_{C_B}
(\mathbb E[D_{B,i}\mid C_B])
\nonumber\\
&=
\frac{F^2}{m_i^2}
\iint_{\mathcal V\times\mathcal V}
K_C(|\mathbf r-\mathbf r'|)
\nonumber\\
&\quad\times
u_i(\mathbf r)u_i(\mathbf r')
\,d^3\mathbf r\,d^3\mathbf r'.
\label{eq:heterogeneity_variance}
\end{align}
The stochastic contribution represents
the mean conditional variance for
a fixed boron field, whereas the
heterogeneity contribution measures
the variation in conditional mean dose
between different field realizations.
Thus, this decomposition applies
to an ensemble that includes both
variations in boron distributions
and repeated capture events.
For repeated irradiation of the
same fixed boron field, only the
conditional variance in
Eq.~\eqref{eq:conditional_variance}
contributes to the fluctuations.

If increasing the neutron fluence
does not change the boron statistics,
neutron energy spectrum, or transport
response, the two variance contributions
scale as $\sigma_{\mathrm{shot},i}^2\propto F$
and $\sigma_{\mathrm{het},i}^2\propto F^2$.
Since the mean dose is proportional
to $F$, the relative dose variance
can be written as
\begin{equation}
\frac{\operatorname{Var}(D_{B,i})}
{\langle D_{B,i}\rangle^2}
=
\frac{A_i}{F}+B_i,
\label{eq:relative_variance}
\end{equation}
where the fluence-independent
coefficients are
\begin{align}
A_i
&=
\frac{
\int_{\mathcal V}v_i(\mathbf r)\,d^3\mathbf r
}{
\overline C_B
\left[
\int_{\mathcal V}u_i(\mathbf r)\,d^3\mathbf r
\right]^2
},
\label{eq:model_relative_shot_coefficient}
\\
B_i
&=
\frac{
\iint_{\mathcal V\times\mathcal V}
K_C(d)u_i(\mathbf r)u_i(\mathbf r')
\,d^3\mathbf r\,d^3\mathbf r'
}{
\overline C_B^{\,2}
\left[
\int_{\mathcal V}u_i(\mathbf r)\,d^3\mathbf r
\right]^2
}.
\label{eq:model_relative_heterogeneity_coefficient}
\end{align}
The term $A_i/F$ decreases with
increasing neutron fluence, whereas
$B_i$ represents the limiting
ensemble relative variance at
high fluence. Its value depends
on the spatial structure of the
boron distribution and the
target response function.
This limiting value does not
imply a persistent relative
variance when irradiation
is repeated for an unchanged
boron distribution.

The relative importance of the
two variance contributions is
measured by
\begin{equation}
\chi_i
=
\frac{\sigma_{\mathrm{het},i}^2}
{\sigma_{\mathrm{shot},i}^2}
=
\frac{F B_i}{A_i}.
\label{eq:crossover}
\end{equation}
Values of $\chi_i<1$ indicate
that stochastic capture fluctuations
dominate, whereas $\chi_i>1$
indicates that boron heterogeneity
is the dominant contribution.
For $B_i>0$, the condition
$\chi_i=1$ defines the crossover
fluence, $\Phi_{n,i}^{*}
=A_i/\sigma_{\mathrm{th}}B_i$. This condition corresponds to equal
contributions from the two variance
components. For a homogeneous boron
field, $B_i=0$, and no finite
crossover fluence exists.

To characterize hit statistics,
the probability that a single
capture event at position
$\mathbf r$ deposits energy
above a threshold is defined as $p_i(\mathbf r)
=
\Pr_\omega
[\varepsilon_i(\mathbf r,\omega)>E_{\min}]$.
In this study, $E_{\min}=0$,
so a hit occurs whenever an
alpha--lithium event deposits
a positive amount of energy
in the target.
Thinning the Poisson process
according to $p_i$ gives
the conditional mean number
of hits \cite{LastPenrose2017},
\begin{equation}
\Lambda_i[C_B]
=
\int_{\mathcal V}
\lambda_C(\mathbf r)p_i(\mathbf r)\,d^3\mathbf r.
\label{eq:mean_hit}
\end{equation}

For a given boron field, the number
of hits is Poisson distributed with
mean $\Lambda_i[C_B]$.
Averaging the conditional zero-hit
probability over boron field
realizations gives
$P_{0,i}=\mathbb{E}_{C_B}
[\exp(-\Lambda_i[C_B])]$. Since the exponential function
is convex, Jensen's inequality gives $P_{0,i}
\geq
\exp\{-\mathbb E_{C_B}[\Lambda_i[C_B]]\}$.
The right-hand side is the
zero-hit probability for a
homogeneous boron distribution
with the same mean concentration,
neutron fluence, and transport
response. This inequality
compares heterogeneous and
homogeneous boron distributions;
the dependence on heterogeneity
amplitude and correlation length
is evaluated for the specified
random-field model.
The quantity $P_{0,i}$
accounts only for hits from
alpha--lithium reaction products
and must not be interpreted
as a cell survival probability.
Such an interpretation requires
additional biological response
modeling, as considered in
previous microdosimetric studies
\cite{Sato2018,Shigehira2026,Yamazaki2026}.

\subsection{Geant4-based numerical methods}
\label{subsec:numerical_method}

Monte Carlo methods have been
widely used in BNCT dosimetry,
including the development of
the COMPASS dose calculation
engine and GATE-based simulations
\cite{Zhong2023,Pistone2025}.
In the present study, the
energy-deposition response from
a single $^{10}$B neutron capture
event is calculated using Geant4
\cite{Agostinelli2003,Allison2016,
Charef2025Geant4BNCT}.
The target is modeled as a
spherical nucleus with radius
$R_n=4\,\mu\mathrm{m}$ embedded
in a water-equivalent medium
with a density of
$1\,\mathrm{g/cm^3}$.
Each history begins with the
generation of an alpha particle
and a $^{7}$Li nucleus at the
same location, emitted in
opposite directions with an
isotropically distributed
orientation. Two reaction
branches are considered, with
probabilities of $93.7\%$
and $6.3\%$ \cite{Sato2018,Han2023}.
The kinetic energies of the
alpha particle and lithium
nucleus are $1.47$ and
$0.84$ MeV, respectively,
for the dominant branch,
and $1.78$ and $1.01$ MeV
for the other branch.
Particle transport is simulated
using
\texttt{G4EmStandardPhysics\_option4}
and \texttt{G4StepLimiterPhysics}
\cite{Geant4PhysicsManual}.
This configuration includes
energy loss and electromagnetic
interactions of charged particles
without enabling inelastic
hadronic nuclear processes.
Neutron capture is modeled
stochastically outside Geant4,
so neutrons are not explicitly
transported. Gamma photons
from the capture reaction,
as well as other neutron and
photon dose components, are
excluded from the calculations.

The response is evaluated at
40 source distances between
$r=0$ and $18\,\mu\mathrm{m}$
from the nucleus center.
At each distance, $10^5$
histories are simulated in
each of five independent runs.
The transport settings use
a maximum step length of
$0.10\,\mu\mathrm{m}$ and a
{production cut} of
$0.05\,\mu\mathrm{m}$.
The latter specifies a
range-based threshold for
secondary particle production,
rather than an energy threshold
for terminating the transport
of an existing particle.
The energy deposited in the
nucleus is recorded for every
history, including those with
zero energy deposition, to
calculate the mean response
$u(r)$, the second raw moment
$v(r)$, and the hit probability
$p(r)$. A hit is defined as
an event that deposits a
positive amount of energy
within the target.
Uncertainties in $u(r)$ and
$v(r)$ are estimated from
the standard errors of the
five independent runs, while
$95\%$ confidence intervals
for $p(r)$ are calculated
using the Wilson method
\cite{Wilson1927}.
The response statistics are
checked against the inequalities
$v(r)\geq u^2(r)$ and
$p(r)v(r)\geq u^2(r)$,
together with the energy
balance of the reaction products.
The radial responses are
linearly interpolated for
spatial integration, while
contributions beyond
$r=18\,\mu\mathrm{m}$ are
neglected in the main calculations.

The boron field is generated
in a cubic domain with side
length $L=40\,\mu\mathrm{m}$
and a baseline resolution of
$80^3$ voxels, corresponding
to a voxel size of
$\Delta x=0.5\,\mu\mathrm{m}$.
A stationary Gaussian field
is generated using a
Fourier-based spectral method
with periodic boundary
conditions and subsequently
transformed into a lognormal
field according to
Eq.~\eqref{eq:boron_field}.
The Gaussian correlation
parameter is chosen so that
the concentration field
after transformation satisfies
the prescribed correlation-length
definition $\xi$.
The mean and variance are
specified at the ensemble
level, without separately
normalizing each field
realization.
The mean boron concentration
is set to $30\,\mu\mathrm{g}$
of $^{10}$B per gram of
medium and converted into
atomic number density using
$\overline C_B=\rho w_{10}N_A/M_{10}$,
where $w_{10}=30\times10^{-6}$,
$N_A$ is Avogadro's constant,
and $M_{10}$ is the molar
mass of $^{10}$B.
The resulting number density
is approximately
$1.8\times10^{18}\,\mathrm{cm^{-3}}$.
Thermal-neutron capture is
calculated using a cross
section of
$\sigma_{\mathrm{th}}=3837$
barn
\cite{Prosdocimi1963BoronCrossSection}.
Changing the neutron fluence
affects only the capture
intensity, while the boron
field statistics and particle
transport responses remain fixed.

The dose variance decomposition
is examined using 2000 independent
boron field realizations, with
eight stochastic repetitions
for each field.
Analytical predictions are
obtained from
Eqs.~\eqref{eq:shot_variance}
and~\eqref{eq:heterogeneity_variance}.
In the MC sampling,
the number of events producing
hits in each volume element
follows a Poisson distribution
conditional on the local
boron concentration and the
response $p(r)$.
The energy deposited by
each hit is sampled from
a bounded surrogate distribution
matched to the Geant4 response
moments. For $p(r)>0$, the
first and second moments of
the deposited energy conditional
on a hit are $u(r)/p(r)$
and $v(r)/p(r)$, respectively,
whereas events without hits
are assigned zero deposited
energy. This construction
preserves the hit probability
and the two energy moments
required to determine the
mean dose and its variance,
but does not reproduce the
complete Geant4 energy-deposition
distribution for individual
histories.

The stochastic variance
contribution is estimated
by averaging the variances
across repetitions within
each fixed boron field.
The heterogeneity contribution
is obtained from the variance
of the field-averaged doses
after correcting for the
stochastic fluctuations
associated with the finite
number of repetitions.
This correction is given
by the estimated stochastic
variance divided by $R=8$.
The estimator is described
in more detail in
Appendix~\ref{app:variance_derivation}.
Accordingly, the comparison
between MC results
and analytical predictions
tests the consistency of
the moment formulation
and the stochastic sampling
procedure, rather than
providing an independent
validation of the full
energy-deposition spectrum
from Geant4.

The nuclear zero-hit probability
is calculated by averaging
$\exp[-\Lambda_i[C_B]]$
over 2000 boron field
realizations and is compared
with Poisson hit-count
sampling using 16 repetitions
per field.
To reduce sampling uncertainty
across field realizations,
a {control variate}
estimator is used, based on
the mean hit intensity
of the homogeneous-boron
reference model \cite{Owen2013}.
Statistical uncertainty is
evaluated using 1000 bootstrap
resamples at the field-realization
level \cite{Efron1979}.
All Poisson repetitions
associated with the same
boron field are retained
as a group during resampling,
rather than treated as
independent field realizations.
Consistency checks include
probability bounds,
Jensen's inequality,
and agreement between
the averaged conditional
Poisson probabilities and
the sampled hit counts.
Since both approaches use
hit intensities obtained
from the same Geant4 response,
their comparison validates
Poisson sampling within the
stochastic model, rather
than providing an independent
Geant4 transport validation.
The probability estimators
and control variate correction
are detailed in
Appendix~\ref{app:zero_hit_derivation}.

Additional numerical checks
include comparisons of Monte
Carlo variances with analytical
predictions, tests of the
Geant4 response cutoff over
$r_{\max}=10$--$18\,\mu\mathrm{m}$,
and spatial grid sensitivity
tests using resolutions of
$64^3$, $80^3$, $100^3$,
and $120^3$ at a fixed
domain size.
As shown in
Appendix~\ref{app:numerical_validation},
the relative difference in
zero-hit probability from
the reference calculation at
$r_{\max}=18\,\mu\mathrm{m}$
decreases to less than
$1\%$ when the response
is retained up to
$r_{\max}=12\,\mu\mathrm{m}$
and becomes very small beyond
$14\,\mu\mathrm{m}$ for the
tested configuration.
The asymptotic relative
variance coefficient and
zero-hit probability also
vary by approximately
$1\%$ or less between
the baseline $80^3$ grid
and the $120^3$ reference
grid, accounting for
sampling uncertainty.
These results support
using the Geant4 response
up to $18\,\mu\mathrm{m}$
and an $80^3$ grid in the
main calculations, although
they do not replace
convergence tests with
respect to domain size
or transport parameters.
The analysis is limited
to a spherical nucleus,
a water-equivalent medium,
a spatially uniform neutron
field, and energy deposition
from alpha--lithium reaction
products within the
electromagnetic transport
model considered.

\section{Results and Discussion}
\label{sec:results}

\subsection{Dose variance decomposition}
\label{subsec:variance_results}

\begin{figure*}[t]
    \centering
    \includegraphics[width=\textwidth]
    {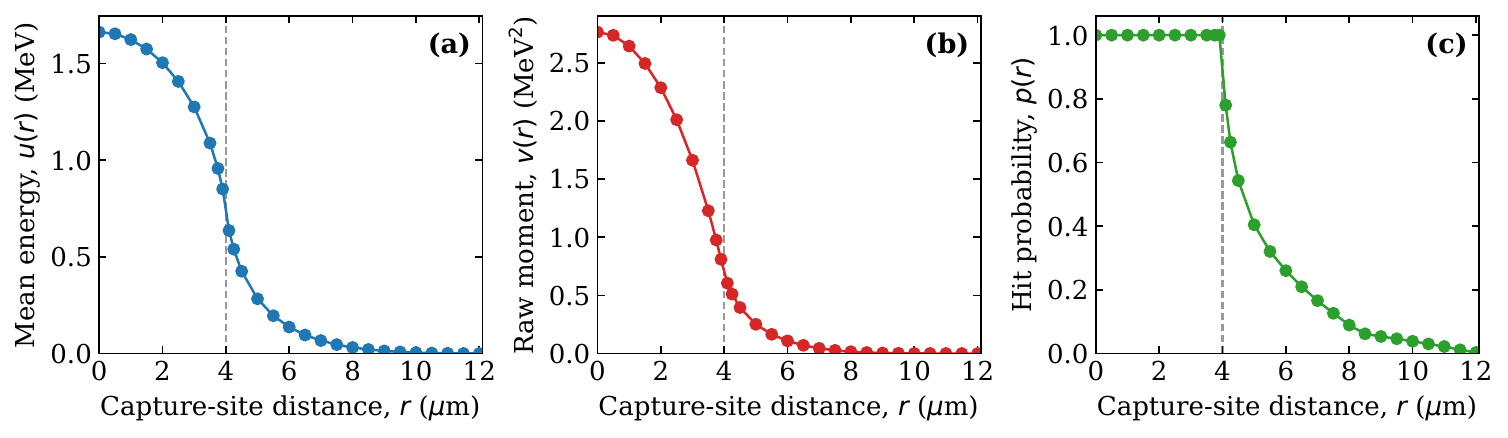}
    \caption{
    Single-event energy-deposition
    responses from $^{10}$B neutron
    capture in a spherical nucleus
    of radius $R_n=4\,\mu\mathrm{m}$
    embedded in a water-equivalent
    medium, calculated using Geant4
    with electromagnetic transport.
    (a) Mean deposited energy,
    $u(r)$; (b) second raw energy
    moment, $v(r)$; and (c) hit
    probability, $p(r)$, as functions
    of the capture-site distance
    from the nucleus center.
    The displayed range is
    $0$--$12\,\mu\mathrm{m}$,
    while the spatial calculations
    use responses extending to
    $18\,\mu\mathrm{m}$.
    The vertical dashed line
    indicates the nuclear boundary
    at $r=R_n$.
    Symbols represent simulation
    results, and lines serve as
    guides to the eye.
    Error bars indicate the standard
    errors from five independent
    runs for $u(r)$ and $v(r)$,
    and $95\%$ Wilson confidence
    intervals for $p(r)$.
    }
    \label{fig:single_event_response}
\end{figure*}

To characterize stochastic energy deposition
at the cellular scale, we first examine the
response from a single $^{10}$B neutron
capture event using Geant4 simulations.
Each capture produces an alpha particle
and a $^{7}$Li nucleus. The energy
deposited in the nucleus depends on the
capture location, emission direction,
and particle trajectories through the
surrounding medium. This dependence on
geometry and particle transport is an
important aspect of BNCT microdosimetry
\cite{SantaCruzZamenhof2004,Han2023}.
The nucleus is modeled as a sphere
with radius $R_n=4\,\mu\mathrm{m}$
embedded in a water-equivalent medium.
The responses are calculated at
40 source distances between $r=0$
and $18\,\mu\mathrm{m}$, while the
range $r\leq12\,\mu\mathrm{m}$ is
shown in Fig.~\ref{fig:single_event_response}
to illustrate their variation near
the nucleus. Three response quantities
are considered: the mean deposited
energy $u(r)$, the second raw energy
moment $v(r)$, and the hit probability
$p(r)$. These quantities are used
to determine the mean dose, dose
variance, and probability that
the nucleus receives no hit,
respectively.

As shown in
Figs.~\ref{fig:single_event_response}(a)
and (b), both the mean deposited
energy and its second raw moment
decrease with increasing distance
between the capture site and the
nucleus center. At $r=0$, the
mean deposited energy is approximately
$1.66$ MeV, while the second raw
moment is about $2.76$ MeV$^2$.
Both quantities decrease as the
source approaches the nuclear
surface and fall more rapidly
for $r>R_n$, where only a
limited range of emission
directions and particle
trajectories can contribute
to energy deposition in the
target. Although $u(r)$ and
$v(r)$ exhibit similar spatial
trends, they describe different
statistical properties.
The mean response $u(r)$
determines the average contribution
of a capture event to the dose,
whereas $v(r)$ is the second
raw moment required to calculate
the variance of a sum of
Poisson-distributed events,
as given by
Eq.~\eqref{eq:shot_variance}.
The variance of the energy
deposited by a single event
at a fixed source location
is $v(r)-u^2(r)$.
However, replacing $v(r)$
with this variance would
omit fluctuations in the
number of neutron capture events.
The importance of target
size in BNCT microdosimetry
has also been demonstrated
experimentally at different
site sizes \cite{Selva2024},
although those measurements
are not used as direct validation
of the present Geant4 responses.

The hit probability in
Fig.~\ref{fig:single_event_response}(c)
illustrates how nuclear geometry
and particle range affect the
likelihood of energy deposition.
For capture sites inside the
nucleus, $p(r)$ is close to
unity and decreases as the
source moves outside the target.
At $r=12\,\mu\mathrm{m}$,
the Geant4 response gives
$p(r)\simeq3.50\times10^{-3}$
and
$u(r)\simeq1.49\times10^{-4}$
MeV. The hit probability
decreases to approximately
$1.2\times10^{-5}$ at
$r=14\,\mu\mathrm{m}$,
while no hits were observed
in the sampled histories
for $r\geq14.5\,\mu\mathrm{m}$.
The absence of observed hits
within the available number
of histories does not imply
that the true hit probability
is exactly zero.
The different behaviors
of $u(r)$ and $p(r)$ also
show that the mean deposited
energy alone cannot determine
whether a capture event
reaches the target.
The response $p(r)$ is
therefore treated separately
when calculating the zero-hit
probability in
Subsection~\ref{subsec:zero_hit_results}.

\begin{figure*}[t]
    \centering
    \includegraphics[width=\textwidth]
    {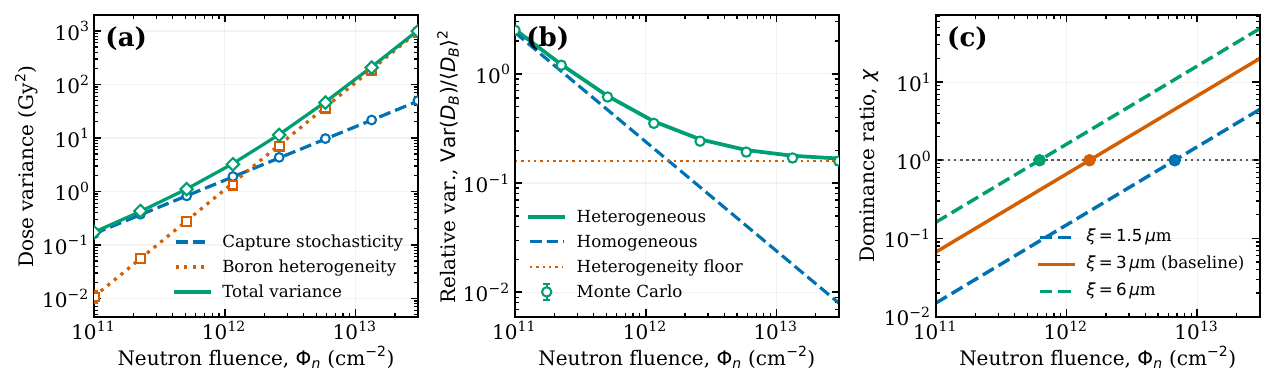}
    \caption{
    Decomposition of nuclear boron
    dose variance as a function
    of neutron fluence $\Phi_n$.
    (a) Contributions from
    stochastic neutron capture
    ($\sigma_{\mathrm{shot}}^2$),
    boron heterogeneity
    ($\sigma_{\mathrm{het}}^2$),
    and the total variance.
    (b) Relative dose variance
    for homogeneous and heterogeneous
    boron distributions, together
    with the high-fluence limit
    arising from boron heterogeneity.
    (c) Dominance ratio
    $\chi=\sigma_{\mathrm{het}}^2/
    \sigma_{\mathrm{shot}}^2$
    for correlation lengths
    $\xi=1.5$, $3$, and
    $6\,\mu\mathrm{m}$.
    The condition $\chi=1$
    marks the crossover fluence
    at which the two variance
    contributions are equal.
    Lines show analytical
    predictions, while symbols
    in (a) and (b) represent
    MC results.
    Symbols in (c) mark the
    analytically determined
    crossover points.
    The calculations use
    Geant4 responses with
    electromagnetic transport,
    $H=0.8$,
    $\bar C_B=30\,\mu\mathrm{g/g}$,
    and $R_n=4\,\mu\mathrm{m}$.
    Monte Carlo checks use
    2000 boron field realizations
    with eight stochastic
    repetitions per field.
    }
    \label{fig:variance_decomposition}
\end{figure*}

These Geant4 responses are then
used to calculate dose statistics
for spatially correlated lognormal
boron fields. According to
Eq.~\eqref{eq:variance_decomposition},
the total dose variance is
separated into the stochastic
capture contribution
$\sigma_{\mathrm{shot}}^2$
and the spatial boron
heterogeneity contribution
$\sigma_{\mathrm{het}}^2$.
The baseline calculations use
a mean $^{10}$B concentration
of $30\,\mu\mathrm{g/g}$,
a heterogeneity amplitude
of $H=0.8$, and a correlation
length of $\xi=3\,\mu\mathrm{m}$.
The neutron fluence is varied
from $10^{11}$ to
$3\times10^{13}\,\mathrm{cm}^{-2}$.
Monte Carlo checks are performed
using 2000 boron field realizations,
with eight stochastic repetitions
per field. The variance decomposition,
relative dose variance, and changes
in the dominant fluctuation
mechanism are presented in
Fig.~\ref{fig:variance_decomposition}.

Figure~\ref{fig:variance_decomposition}(a)
shows that the two variance
contributions have different
dependencies on neutron fluence.
The stochastic contribution
increases linearly,
$\sigma_{\mathrm{shot}}^2\propto\Phi_n$,
whereas the heterogeneity
contribution increases
quadratically,
$\sigma_{\mathrm{het}}^2\propto\Phi_n^2$.
At low fluence, the total
variance is mainly determined
by fluctuations in the number
of capture events and the
energy deposited by each event.
As fluence increases, the
heterogeneity contribution
grows more rapidly and
eventually becomes dominant.
For the baseline configuration,
the two contributions become
equal at
$\Phi_n^*\simeq1.50\times10^{12}
\,\mathrm{cm}^{-2}$.
This crossover identifies
a change in the dominant
source of dose fluctuations
within the model, rather
than a phase transition
or a threshold for biological
effectiveness. The MC
results follow the analytical
predictions over the fluence
range examined.

The difference between the
two mechanisms becomes clearer
when the variance is normalized
by the square of the mean dose,
as shown in
Fig.~\ref{fig:variance_decomposition}(b).
For a homogeneous boron distribution,
the relative variance decreases
inversely with neutron fluence,
because only stochastic capture
fluctuations contribute.
For heterogeneous boron fields,
the relative variance instead
approaches a constant at
high fluence as the
heterogeneity contribution
becomes dominant.
For the baseline parameters,
Eq.~\eqref{eq:relative_variance}
gives the numerical relation
\begin{equation}
\frac{\operatorname{Var}(D_B)}
{\langle D_B\rangle^2}
\simeq
\frac{2.39\times10^{11}\,
\mathrm{cm}^{-2}}{\Phi_n}
+0.160.
\label{eq:relative_variance_numerical}
\end{equation}
The first term represents
the relative fluctuations
arising from stochastic
neutron capture, which
decrease with increasing
fluence. The second term
determines the limiting
relative variance at high
fluence. The plateau value
$B\simeq0.160$ depends on
the heterogeneity amplitude,
the spatial correlation
structure of the boron field,
and the target response,
and is therefore not universal.
Importantly, this limit
describes variations between
boron field realizations,
rather than a persistent
relative variance under
repeated irradiation of
the same fixed distribution.
The importance of boron
microdistribution and
stochastic energy deposition
has been discussed in
previous microdosimetric
models \cite{Sato2018,Marcaccio2026}.
Here, we specifically examine
how the relative contributions
of these two variance
mechanisms change with
neutron fluence and boron
correlation length, without
modeling biological responses.

The influence of correlation
length on the crossover
is shown in
Fig.~\ref{fig:variance_decomposition}(c).
The dominance ratio
$\chi=\sigma_{\mathrm{het}}^2/
\sigma_{\mathrm{shot}}^2$
increases linearly with
neutron fluence, but the
position of $\chi=1$ depends
on the boron correlation length.
For $H=0.8$, the crossover
fluences are
$6.72\times10^{12}$,
$1.50\times10^{12}$, and
$6.21\times10^{11}\,
\mathrm{cm}^{-2}$
for $\xi=1.5$, $3$, and
$6\,\mu\mathrm{m}$,
respectively. Increasing
the correlation length
from $1.5$ to $6\,\mu\mathrm{m}$
therefore reduces the crossover
fluence by approximately
one order of magnitude.
This behavior arises because
boron concentration variations
correlated over longer
distances undergo less
spatial averaging through
the energy-deposition response.
As a result, the differences
in conditional mean dose
between field realizations
become larger, allowing
boron heterogeneity to
dominate at lower fluence.
The crossover points in
panel (c) are obtained
analytically rather than
from independent Monte
Carlo estimates.

Quantitative agreement
between MC sampling
and the analytical predictions
is assessed using the ratio
of the two results for
each variance component.
With 2000 boron field
realizations, the Monte
Carlo-to-theory ratios range
from $0.972$ to $1.028$
for the stochastic contribution,
$0.980$ to $1.101$ for
the heterogeneity contribution,
and $0.995$ to $1.034$
for the total variance.
As a separate check,
the variance of the conditional
mean doses across field
realizations gives a ratio
of $1.020$ relative to
the analytical heterogeneity
prediction. These results
show that the variance
decomposition is reproduced
consistently by stochastic
sampling, although the
heterogeneity component has
a larger statistical uncertainty.
Additional sensitivity tests
involving the Geant4 response
cutoff and spatial grid
resolution are presented
in Appendix~
\ref{app:numerical_validation}.

These results
show that nuclear dose
fluctuations arise from
two mechanisms with distinct
fluence dependencies.
The boron correlation length
controls the magnitude
of the heterogeneity
contribution and the fluence
at which it becomes dominant,
even when the mean boron
concentration remains fixed.
Thus, mean absorbed dose
alone does not adequately
characterize dose variations
at the nuclear scale.
These findings are limited
to alpha--lithium energy
deposition, the target
geometry, and the statistical
properties of the boron
fields considered here.

\subsection{Nuclear zero-hit probability}
\label{subsec:zero_hit_results}

\begin{figure*}[tb!]
    \centering
    \includegraphics[width=\textwidth]
    {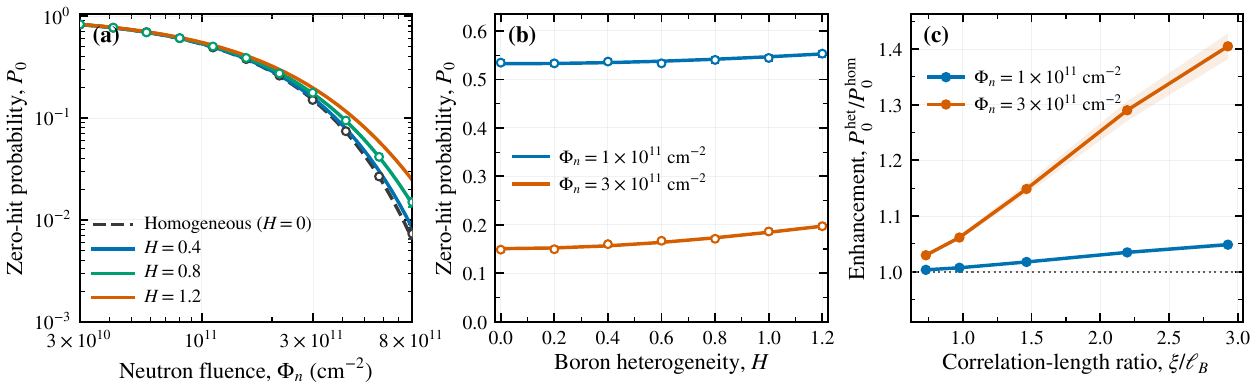}
    \caption{
    Nuclear zero-hit probability
    ($P_0$) associated with
    alpha--lithium reaction products
    in heterogeneous boron fields.
    (a) Dependence of $P_0$ on
    neutron fluence $\Phi_n$
    for different heterogeneity
    amplitudes $H$ at
    $\xi=3\,\mu\mathrm{m}$.
    (b) Variation of $P_0$
    with $H$ at
    $\Phi_n=1\times10^{11}$
    and $3\times10^{11}\,
    \mathrm{cm}^{-2}$,
    with $\xi=3\,\mu\mathrm{m}$.
    (c) Zero-hit probability ratio
    $P_0^{\mathrm{het}}/
    P_0^{\mathrm{hom}}$
    as a function of the
    normalized correlation length
    $\xi/\ell_B$, where
    $\ell_B=2.05\,\mu\mathrm{m}$
    is an external reference
    length scale and $H=0.8$.
    Curves represent the ensemble
    averages of the conditional
    Poisson zero-hit probabilities,
    while open symbols in
    (a) and (b) show MC
    sampling of Poisson hit counts.
    Shaded bands in (c) indicate
    $95\%$ bootstrap confidence
    intervals. The calculations
    use Geant4 responses with
    electromagnetic transport,
    $\bar C_B=30\,\mu\mathrm{g/g}$,
    $R_n=4\,\mu\mathrm{m}$,
    and 2000 boron field
    realizations with 16 Poisson
    repetitions per field.
    }
    \label{fig:zero_hit_probability}
\end{figure*}

In addition to affecting dose
variance, spatial boron heterogeneity
also influences the probability
that a nucleus receives no energy
deposition from neutron capture
reaction products. The zero-hit
probability, $P_0$, is defined
as the probability that no event
deposits a positive amount of
energy within the nucleus during
irradiation. In the Poisson model,
this probability depends on the
neutron capture intensity and
the single-event hit probability
$p(r)$ obtained from the Geant4
simulations in
Fig.~\ref{fig:single_event_response}(c).
For a given boron field realization,
the zero-hit probability is
$\exp[-\Lambda[C_B]]$, while
the ensemble probability is
obtained by averaging this
quantity over the correlated
lognormal boron fields.
The calculations use Geant4
responses with electromagnetic
transport, a mean $^{10}$B
concentration of
$30\,\mu\mathrm{g/g}$, and
a spherical nucleus of radius
$R_n=4\,\mu\mathrm{m}$.
The effects of neutron fluence,
heterogeneity amplitude $H$,
and correlation length $\xi$
on the zero-hit probability
are presented in
Fig.~\ref{fig:zero_hit_probability}.

Figure~\ref{fig:zero_hit_probability}(a)
shows that the zero-hit probability
decreases monotonically with
increasing neutron fluence.
This decrease occurs because
higher fluence increases the
expected number of neutron
captures capable of depositing
energy within the nucleus.
At low fluence, the curves
for homogeneous and heterogeneous
boron distributions nearly
overlap, whereas their differences
become more pronounced as
fluence increases. At a given
fluence, larger heterogeneity
amplitudes produce higher
zero-hit probabilities over
the parameter range examined.
Although all zero-hit
probabilities decrease at
high fluence, differences
between homogeneous and
heterogeneous fields remain
visible on a relative scale.
Thus, increasing neutron
fluence reduces the likelihood
of receiving no hit, but
does not eliminate the
influence of spatial boron
heterogeneity on hit statistics.

The effect of heterogeneity
amplitude is shown more clearly
in Fig.~\ref{fig:zero_hit_probability}(b),
which compares $P_0$ at two
neutron fluences for a fixed
correlation length of
$\xi=3\,\mu\mathrm{m}$.
For $H=0.8$ and
$\Phi_n=1\times10^{11}\,
\mathrm{cm}^{-2}$, the
zero-hit probability increases
from $0.5323$ for homogeneous
boron to $0.5416$ for the
heterogeneous distribution,
corresponding to a relative
increase of approximately
$1.75\%$. At
$\Phi_n=3\times10^{11}\,
\mathrm{cm}^{-2}$, the
probability increases from
$0.1508$ to $0.1732$,
equivalent to a relative
increase of $14.87\%$.
These results show that
the influence of heterogeneity
on zero-hit probability
becomes more pronounced at
higher fluence within the
range studied, even though
the absolute probabilities
are smaller.

Variations in boron concentration
broaden the distribution of hit
intensities across field realizations.
Realizations with lower hit intensities
make a relatively larger contribution
to the ensemble zero-hit probability
because of the exponential dependence
in the Poisson model.
This behavior is consistent with
Jensen's inequality,
$P_0^{\mathrm{het}}\geq P_0^{\mathrm{hom}}$,
for the same mean boron concentration,
neutron fluence, and transport response.
However, this inequality does not
by itself guarantee that $P_0$
increases monotonically with every
heterogeneity parameter.
The trends observed in
Fig.~\ref{fig:zero_hit_probability}(b)
are specific to the random-field
model and parameter range
considered here.

In addition to heterogeneity
amplitude, the boron correlation
length has a substantial effect
on the zero-hit probability,
as shown in
Fig.~\ref{fig:zero_hit_probability}(c).
The correlation length is
expressed as $\xi/\ell_B$,
where $\ell_B=2.05\,\mu\mathrm{m}$
is a reference length scale
adopted from an earlier
energy-deposition response study.
This value is not derived
from the hit probability
$p(r)$ in
Fig.~\ref{fig:single_event_response}
and is used only to normalize
the horizontal axis, rather
than to define a universal
physical threshold.

For $H=0.8$ and
$\Phi_n=3\times10^{11}\,
\mathrm{cm}^{-2}$,
the ratio
$P_0^{\mathrm{het}}/
P_0^{\mathrm{hom}}$
increases from $1.0295$
at $\xi=1.5\,\mu\mathrm{m}$
to $1.1487$ at
$\xi=3\,\mu\mathrm{m}$,
reaching $1.4049$ at
$\xi=6\,\mu\mathrm{m}$.
These values correspond
to relative increases in
zero-hit probability of
$2.95\%$, $14.87\%$,
and $40.49\%$, respectively.
For $\xi=6\,\mu\mathrm{m}$,
the $95\%$ bootstrap
confidence interval for
the relative increase is
$38.20$--$42.81\%$.
At the lower fluence of
$\Phi_n=1\times10^{11}\,
\mathrm{cm}^{-2}$,
the effect of correlation
length remains visible,
but the relative enhancement
is considerably smaller.
These findings show that
heterogeneity amplitude
alone is insufficient
to characterize hit statistics,
because the spatial extent
of regions with different
boron concentrations also
affects the distribution
of effective hit intensities
within the nucleus.

The role of correlation
length can be understood
in terms of spatial averaging
through the hit response.
When $\xi$ is relatively
small, the effective hit
intensity receives contributions
from many regions with
different boron concentrations,
so local concentration
fluctuations tend to average out.
At larger correlation lengths,
boron-rich and boron-poor
regions extend over greater
distances, and neighboring
concentration variations
become more strongly correlated.
This increases fluctuations
in hit intensity across
field realizations and,
through the exponential
dependence of the Poisson
zero-hit probability, raises
the ensemble-averaged $P_0$.
This mechanism is consistent
with the dose variance results
in Fig.~\ref{fig:variance_decomposition},
where longer correlation
lengths also increase the
heterogeneity contribution
and shift the crossover
fluence to lower values.

The numerical consistency
of the zero-hit calculations
is examined by comparing
the ensemble average of
the conditional Poisson
probabilities with direct
Poisson hit-count sampling
on the same boron fields.
Using 2000 field realizations
and 16 repetitions per field,
the maximum absolute difference
between the two estimates
is approximately $0.0057$.
Confidence intervals are
obtained by bootstrap
resampling at the field
realization level.
The agreement supports
the implementation of
the stochastic model,
but does not constitute
an independent validation
of Geant4 particle transport.
The hit response $p(r)$
is taken from Geant4
calculations extending to
$r=18\,\mu\mathrm{m}$.
Sensitivity to the response
cutoff and spatial grid
resolution is examined
separately in
Appendix~\ref{app:numerical_validation}.
These tests support the
numerical stability of the
results for the configurations
considered, although the
effects of domain size and
transport parameters remain
limitations of the analysis.

The results show
that the nuclear zero-hit
probability depends not
only on neutron fluence
and mean boron concentration,
but also on the amplitude
and correlation length
of boron heterogeneity.
Boron fields with the
same mean concentration
can produce different
zero-hit probabilities
because their distributions
of capture intensity
at the micrometer scale
are not identical.
This result complements
the dose variance analysis
by quantifying the probability
that alpha--lithium reaction
products deposit no energy
within the nucleus.
However, the calculated
$P_0$ does not represent
a cell survival probability,
because the model does
not include all BNCT dose
components or cellular
biological responses.
Predicting cell survival
requires additional
biological modeling,
as considered in previous
microdosimetric studies
\cite{Sato2018,Shigehira2026,Yamazaki2026}.

\section{Conclusions}
\label{sec:conclusions}

This study investigated how
stochastic neutron capture
and spatial boron heterogeneity
affect nuclear dose statistics
in boron neutron capture therapy
(BNCT). By combining Geant4
energy-deposition responses,
a Poisson capture process,
and spatially correlated
lognormal boron fields,
the dose variance was separated
into contributions from
stochastic capture fluctuations
and boron heterogeneity.
These contributions increase
linearly and quadratically
with neutron fluence,
respectively, leading to
a correlation-length-dependent
crossover between the dominant
sources of dose fluctuations.
For $H=0.8$ and
$\xi=3\,\mu\mathrm{m}$,
the crossover occurs at
approximately
$1.50\times10^{12}\,
\mathrm{cm}^{-2}$.
Increasing the correlation
length from $1.5$ to
$6\,\mu\mathrm{m}$ reduces
the crossover fluence from
$6.72\times10^{12}$ to
$6.21\times10^{11}\,
\mathrm{cm}^{-2}$.
When normalized by the
square of the mean dose,
the stochastic contribution
decreases with fluence,
whereas boron heterogeneity
produces a limiting relative
variance of $0.160$ for
the baseline configuration.

Boron heterogeneity also
increases the probability
that a nucleus receives
no hit from alpha--lithium
reaction products, even
when the mean boron
concentration remains unchanged.
For $H=0.8$,
$\xi=6\,\mu\mathrm{m}$,
and $\Phi_n=3\times10^{11}\,
\mathrm{cm}^{-2}$,
the zero-hit probability
is $40.49\%$ higher than
that for the homogeneous
boron reference, with a
$95\%$ bootstrap confidence
interval of $38.20$--$42.81\%$
for the relative increase.
These findings show that
both the amplitude and
correlation length of
boron heterogeneity affect
not only the magnitude
of dose fluctuations,
but also the probability
of no energy deposition
within the nucleus.
Thus, mean boron concentration
alone is insufficient
to characterize dose
statistics at the
micrometer scale.

This study provides a quantitative
framework for understanding how
spatial boron heterogeneity affects
nuclear dose fluctuations beyond
conventional mean-dose-based dosimetry.
The agreement between analytical
predictions and MC sampling,
supported by numerical sensitivity
tests, demonstrates the consistency
of the proposed approach.
By explicitly accounting for
stochastic neutron capture and
spatial boron correlations, the
framework reveals how these
mechanisms influence dose variance
and zero-hit probability at the
cellular scale. Although the model
focuses on energy deposition from
alpha--lithium reaction products,
the findings provide a useful
basis for assessing the effects
of boron microdistribution and
the limitations of homogeneous-boron
assumptions in BNCT dosimetry.

\section*{Author Contributions}
\textbf{R.~Fardela:} Writing,  Formal analysis, Validation, Review \& editing. \textbf{R.~Muslim:} Conceptualization,  Methodology, Writing, Software, Formal analysis, Validation, Visualisation, Review \& editing. \textbf{T.~A.~I.~Kusuma:} Writing, Formal analysis, Validation, Review \& editing.  All authors read and reviewed the paper.
\section*{Declaration of Interests}
The contributors declare that they have no apparent competing business or personal connections that might have appeared to have influenced the reported work.
\section*{Acknowledgments}
\textbf{R.~Muslim} was supported by the YST Program of the Asia Pacific Center for Theoretical Physics (APCTP), funded by the Science and Technology Promotion Fund and Lottery Fund of the Korean Government, and by the Management Talent Program of the National Research and Innovation Agency of Indonesia (BRIN).

\section*{Appendix}
\appendix

\section{Boron dose variance decomposition}
\label{app:variance_derivation}

For the lognormal boron field defined in
Eq.~\eqref{eq:boron_field}, let
$d=|\mathbf r-\mathbf r'|$ and
$\rho_X(d)=\exp[-d^2/(2a^2)]$.
Using the joint moment properties
of Gaussian random fields, we obtain
\begin{align}
\langle C_B(\mathbf r)C_B(\mathbf r')\rangle
&=
\overline C_B^{\,2}
\exp[s^2\rho_X(d)],
\\
K_C(d)
&=
\overline C_B^{\,2}
\left\{\exp[s^2\rho_X(d)]-1\right\}.
\label{eq:app_boron_covariance}
\end{align}
Here, $K_C(d)$ denotes the covariance
of boron concentrations at two positions.
Since $s^2=\ln(1+H^2)$, the normalized
boron concentration correlation function
for $H>0$ is
\begin{equation}
\rho_B(d)
=
\frac{\exp[s^2\rho_X(d)]-1}{H^2}.
\label{eq:app_boron_correlation}
\end{equation}
Defining the boron correlation length
through $\rho_B(\xi)=\exp(-1/2)$,
the Gaussian correlation parameter
is determined by
\begin{equation}
\begin{aligned}
\eta(H)
&=
\frac{\ln(1+H^2e^{-1/2})}
{\ln(1+H^2)}, \\
a
&=
\frac{\xi}{\sqrt{-2\ln\eta(H)}}.
\label{eq:app_correlation_calibration}
\end{aligned}
\end{equation}
In the limit $H\to0$, we have
$\eta(H)\to e^{-1/2}$ and $a\to\xi$.
For $H=0$, the boron field becomes
homogeneous, so the normalized
correlation function is no longer
defined and the correlation length
does not affect the results.

To derive the dose statistics,
the total deposited energy $E_i$
in Eq.~\eqref{eq:cellular_dose}
is treated as a sum of marks
associated with a Poisson process.
Assuming that individual capture
histories are conditionally independent
for a given boron field, the conditional
cumulant-generating function is
\begin{align}
K_{E_i}(t\mid C_B)
&=
\ln\mathbb E
\left[e^{tE_i}\mid C_B\right]
\nonumber\\
&=
\int_{\mathcal V}
\lambda_C(\mathbf r)
\left[
\mathbb E_\omega
\left(e^{t\varepsilon_i(\mathbf r,\omega)}\right)
-1
\right]d^3\mathbf r.
\label{eq:app_cumulant_generator}
\end{align}
Differentiating with respect to $t$
at $t=0$ gives the first two
conditional cumulants,
\begin{align}
K_{E_i}'(0\mid C_B)
&=
\int_{\mathcal V}
\lambda_C(\mathbf r)u_i(\mathbf r)
\,d^3\mathbf r,
\\
K_{E_i}''(0\mid C_B)
&=
\int_{\mathcal V}
\lambda_C(\mathbf r)v_i(\mathbf r)
\,d^3\mathbf r.
\label{eq:app_event_cumulants}
\end{align}
The first cumulant is the conditional
mean deposited energy, while the
second is its conditional variance.
Thus, the variance of the total
deposited energy depends on the
second raw moment $v_i$, rather
than the single-event energy
variance $v_i-u_i^2$.
This distinction arises because
the Poisson process accounts for
fluctuations in both the number
of capture events and the energy
deposited by each event.

Converting the deposited energy
into absorbed dose through
$D_{B,i}=E_i/m_i$ and applying
the law of total variance yields
\begin{align}
\operatorname{Var}(D_{B,i})
&=
\mathbb E_{C_B}
\left[
\operatorname{Var}(D_{B,i}\mid C_B)
\right]
\nonumber\\
&\quad+
\operatorname{Var}_{C_B}
\left(
\mathbb E[D_{B,i}\mid C_B]
\right).
\label{eq:app_total_variance_law}
\end{align}
To make the dependence on the
capture factor $F$ explicit,
we define three coefficients,
\begin{align}
\mathcal U_i
&=
\frac{\overline C_B}{m_i}
\int_{\mathcal V}
u_i(\mathbf r)\,d^3\mathbf r,
\\
\mathcal V_i
&=
\frac{\overline C_B}{m_i^2}
\int_{\mathcal V}
v_i(\mathbf r)\,d^3\mathbf r,
\\
\mathcal W_i
&=
\frac{1}{m_i^2}
\iint_{\mathcal V\times\mathcal V}
K_C(|\mathbf r-\mathbf r'|)
\nonumber\\
&\quad\times
u_i(\mathbf r)u_i(\mathbf r')
\,d^3\mathbf r\,d^3\mathbf r'.
\label{eq:app_dose_coefficients}
\end{align}
Substituting
$\lambda_C(\mathbf r)=FC_B(\mathbf r)$
into the total variance relation gives
\begin{equation}
\langle D_{B,i}\rangle=F\mathcal U_i,
\qquad
\sigma_{\mathrm{shot},i}^2=F\mathcal V_i,
\qquad
\sigma_{\mathrm{het},i}^2=F^2\mathcal W_i.
\label{eq:app_fluence_scaling}
\end{equation}
These coefficients are independent
of neutron fluence as long as the
boron field statistics and transport
responses remain unchanged.
The relative dose variance and
the dominance ratio are therefore
\begin{align}
\frac{\operatorname{Var}(D_{B,i})}
{\langle D_{B,i}\rangle^2}
&=
\frac{\mathcal V_i}
{F\mathcal U_i^2}
+
\frac{\mathcal W_i}{\mathcal U_i^2},
\\
\chi_i
&=
\frac{F\mathcal W_i}{\mathcal V_i}.
\label{eq:app_relative_variance_and_chi}
\end{align}
Under the thermal-neutron
approximation,
$F=\sigma_{\mathrm{th}}\Phi_n$.
The condition $\chi_i=1$
then defines the crossover
fluence,
\begin{equation}
\Phi_{n,i}^{*}
=
\frac{\mathcal V_i}
{\sigma_{\mathrm{th}}\mathcal W_i},
\label{eq:app_crossover_fluence}
\end{equation}
provided that $\mathcal W_i>0$.
For a homogeneous boron distribution,
$K_C(d)=0$ and $\mathcal W_i=0$,
so no finite crossover fluence exists.
These relations also show that the
high-fluence limit of the relative
dose variance is determined by
$\mathcal W_i/\mathcal U_i^2$,
which represents variations in the
conditional mean dose across boron
field realizations.

For the MC consistency
checks, let $D_{j,k}$ denote the
absorbed dose for boron field
realization $j$ and stochastic
repetition $k$. Let
$\overline D_j$ be the mean
dose over $R$ repetitions
for field $j$, and let $s_j^2$
be the corresponding sample
variance. The two variance
contributions are estimated as
\begin{align}
\widehat{\sigma}_{\mathrm{shot}}^2
&=
\frac{1}{M}
\sum_{j=1}^{M}s_j^2,
\\
\widehat{\sigma}_{\mathrm{het}}^2
&=
s_{\overline D}^{\,2}
-
\frac{\widehat{\sigma}_{\mathrm{shot}}^2}{R},
\label{eq:app_mc_variance_estimator}
\end{align}
where $s_{\overline D}^{\,2}$
is the sample variance of the
field-averaged doses
$\overline D_j$ across the
$M$ field realizations.
The second term is subtracted
because the mean over a finite
number of repetitions still
contains stochastic capture
fluctuations.
For independent field realizations
and unbiased sample variance
estimates, this procedure
separates the two contributions
in expectation.

\section{Zero-hit probability estimation}
\label{app:zero_hit_derivation}

For a fixed boron field, each neutron
capture at position $\mathbf r$
produces a hit in the nucleus with
probability $p_i(\mathbf r)$.
By the thinning property of
Poisson processes, the conditional
number of hits also follows a
Poisson distribution with mean
$\Lambda_i[C_B]$.
Its probability-generating function is
\begin{equation}
G_i(z\mid C_B)
=
\exp\left[
(z-1)\Lambda_i[C_B]
\right],
\label{eq:app_hit_pgf}
\end{equation}
where $\Lambda_i[C_B]$ is defined
in Eq.~\eqref{eq:mean_hit}.
Setting $z=0$ and averaging over
the ensemble of boron fields gives
\begin{equation}
P_{0,i}
=
\mathbb E_{C_B}
\left[
e^{-\Lambda_i[C_B]}
\right].
\label{eq:app_zero_hit_expectation}
\end{equation}
For thermal neutrons, the mean
hit intensity can be expressed as
$\Lambda_i[C_B]=\Phi_nQ_i[C_B]$,
where
\begin{align}
Q_i[C_B]
&=
\sigma_{\mathrm{th}}
\int_{\mathcal V}
C_B(\mathbf r)p_i(\mathbf r)
\,d^3\mathbf r,
\\
Q_{0,i}
&=
\sigma_{\mathrm{th}}\overline C_B
\int_{\mathcal V}
p_i(\mathbf r)\,d^3\mathbf r.
\label{eq:app_hit_rate_coefficients}
\end{align}
Since
$\mathbb E_{C_B}[Q_i]=Q_{0,i}$,
the zero-hit probability for a
homogeneous boron distribution
with the same mean concentration
is
$P_{0,i}^{\mathrm{hom}}
=\exp(-\Phi_nQ_{0,i})$.
It follows that
\begin{equation}
\frac{P_{0,i}^{\mathrm{het}}}
{P_{0,i}^{\mathrm{hom}}}
=
\mathbb E_{C_B}
\left[
e^{-\Phi_n(Q_i-Q_{0,i})}
\right]
\geq 1.
\label{eq:app_jensen_ratio}
\end{equation}
This inequality follows from the
convexity of the exponential
function and Jensen's inequality.
It guarantees that the ensemble
zero-hit probability for a
heterogeneous boron distribution
is no smaller than that of the
homogeneous reference with the
same mean hit intensity.
However, it does not generally
guarantee a monotonic dependence
on every heterogeneity parameter.
The effects of $H$ and $\xi$
depend on the statistical
properties of the boron field
considered.

The zero-hit probability is
evaluated using $M$ independent
boron field realizations with
hit-intensity coefficients
$Q_{i,j}$.
To reduce sampling uncertainty
across fields, we use a
{control variate} estimator
based on the analytically known
mean intensity $Q_{0,i}$
\cite{Owen2013},
\begin{align}
\widehat P_{0,i}^{\mathrm{CV}}
&=
\frac{1}{M}
\sum_{j=1}^{M}
\Bigl[
e^{-\Phi_nQ_{i,j}}
\nonumber\\
&\quad+
\Phi_n e^{-\Phi_nQ_{0,i}}
(Q_{i,j}-Q_{0,i})
\Bigr].
\label{eq:app_cv_estimator}
\end{align}
The correction term is based
on a linear approximation
to the fluctuations in
$Q_{i,j}-Q_{0,i}$.
It has zero expectation because
$\mathbb E[Q_{i,j}]=Q_{0,i}$.
The estimator therefore remains
unbiased for the ensemble
zero-hit probability $P_{0,i}$,
while reducing the variation
associated with fluctuations
in the hit intensity.

As a separate consistency check
of Poisson sampling, hit counts
are generated according to
$K_{j,k}\sim
\operatorname{Poisson}(\Phi_nQ_{i,j})$
for repetitions $k=1,\ldots,R$.
Using the same control variate,
the MC estimator is
defined as
\begin{align}
\widehat P_{0,i}^{\mathrm{MC}}
&=
\frac{1}{M}
\sum_{j=1}^{M}
\Biggl[
\frac{1}{R}
\sum_{k=1}^{R}
\mathbf 1(K_{j,k}=0)
\nonumber\\
&\quad+
\Phi_n e^{-\Phi_nQ_{0,i}}
(Q_{i,j}-Q_{0,i})
\Biggr].
\label{eq:app_poisson_mc_estimator}
\end{align}
Because
$\mathbb E[\mathbf 1(K_{j,k}=0)
\mid Q_{i,j}]
=e^{-\Phi_nQ_{i,j}}$,
the conditional expectation
of the MC estimator,
for a given set of boron fields,
equals the corresponding
conditional Poisson estimator.
Both approaches therefore have
the same ensemble expectation,
although their sampling
uncertainties differ.
This comparison tests the
consistency of Poisson hit-count
sampling and does not constitute
an independent Geant4 transport
simulation.

The $95\%$ confidence intervals
are estimated by bootstrap
resampling \cite{Efron1979}
at the field-realization level.
All Poisson repetitions belonging
to the same boron field are
retained as a group, preserving
the dependence between outcomes
generated from that field.
The main zero-hit calculations
use $M=2000$ boron field
realizations and $R=16$
Poisson repetitions per field,
with $1000$ bootstrap resamples
used to estimate statistical
uncertainty.

\section{Numerical validation}
\label{app:numerical_validation}

\begin{figure*}[tb!]
    \centering
    \includegraphics[width=\textwidth]
    {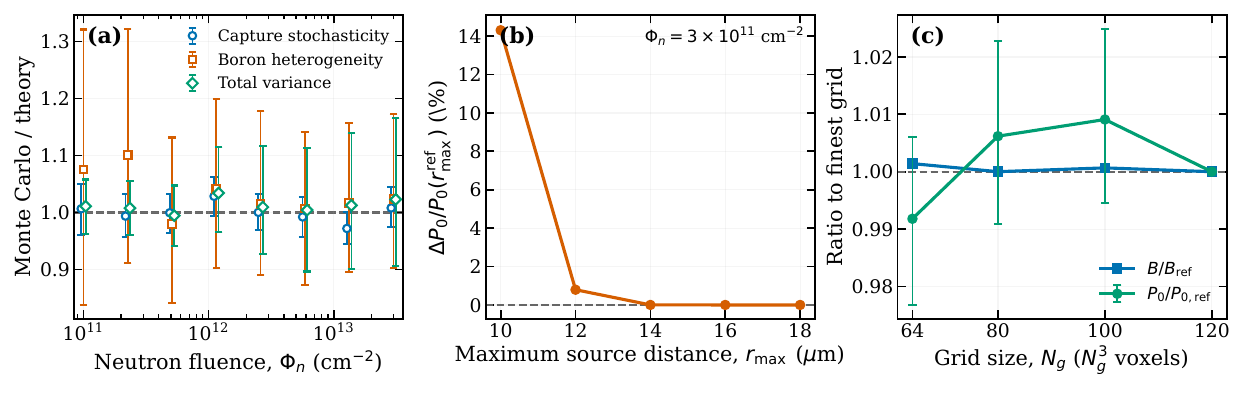}
    \caption{
    Numerical validation and
    sensitivity analysis of the
    stochastic boron dose model.
    (a) Ratios of MC
    variances to analytical
    predictions for stochastic
    capture fluctuations,
    boron heterogeneity, and
    total dose variance as
    functions of neutron fluence.
    (b) Relative difference
    in zero-hit probability
    $P_0$ as a function of
    the maximum source distance
    $r_{\max}$, using
    $r_{\max}=18\,\mu\mathrm{m}$
    as the reference.
    (c) Sensitivity of the
    asymptotic relative variance
    coefficient $B$ and the
    zero-hit probability $P_0$
    to spatial grid resolution,
    normalized by the results
    obtained with the $120^3$ grid.
    Calculations use Geant4
    responses with electromagnetic
    transport,
    $H=0.8$,
    $\xi=3\,\mu\mathrm{m}$,
    $\bar C_B=30\,\mu\mathrm{g/g}$,
    and $R_n=4\,\mu\mathrm{m}$.
    Panels (b) and (c) use
    $\Phi_n=3\times10^{11}\,
    \mathrm{cm}^{-2}$.
    Error bars in (a) and (c)
    represent $95\%$ bootstrap
    confidence intervals.
    }
    \label{fig:numerical_validation}
\end{figure*}

The reliability of the numerical
calculations is assessed by comparing
MC results with analytical
predictions and examining the
sensitivity to the Geant4 response
cutoff and spatial grid resolution,
as shown in
Fig.~\ref{fig:numerical_validation}.
In Fig.~\ref{fig:numerical_validation}(a),
the MC-to-theory variance
ratios range from $0.972$ to $1.028$
for the stochastic contribution,
$0.980$ to $1.101$ for the
heterogeneity contribution, and
$0.995$ to $1.034$ for the
total variance.
These comparisons use 2000 boron
field realizations with eight
stochastic repetitions per field.
The maximum deviation in the
total variance is approximately
$3.4\%$, while the heterogeneity
component shows greater statistical
variation. These results support
the consistency of the variance
decomposition at the level of
the response moments used
in the model.

The effect of truncating the
radial response is shown in
Fig.~\ref{fig:numerical_validation}(b).
Relative to the calculation
using responses up to
$r_{\max}=18\,\mu\mathrm{m}$,
the difference in $P_0$
is approximately $14\%$
when the response is truncated
at $10\,\mu\mathrm{m}$.
This difference decreases
to less than $1\%$ at
$12\,\mu\mathrm{m}$ and
becomes very small when the
cutoff is extended to
$14\,\mu\mathrm{m}$.
These results indicate that
the contribution from the
response tail becomes small
for the configuration examined,
although they do not establish
that the physical hit probability
is exactly zero beyond the cutoff.

Figure~\ref{fig:numerical_validation}(c)
shows the sensitivity to spatial
grid resolutions of $64^3$,
$80^3$, $100^3$, and $120^3$
within the same computational domain.
The coefficient $B$ changes only
slightly, while the difference
in $P_0$ between the baseline
$80^3$ grid and the reference
$120^3$ grid is on the order
of $1\%$ or less, with some
variation attributable to
field-sampling uncertainty.
The sensitivity tests in
panels (b) and (c) use
1000 boron field realizations. These results support
the use of Geant4 responses
extending to $18\,\mu\mathrm{m}$
and an $80^3$ spatial grid
for the main calculations
over the parameter range examined.

\printcredits

\bibliographystyle{cas-model2-names}

\bibliography{cas-refs}





\end{document}